\documentclass{mem}
\usepackage{euclid}
\usepackage{natbib}
\usepackage{txfonts}
\usepackage{balance}
\usepackage{flushend}
\usepackage{graphicx}
\usepackage[a4paper,breaklinks,pdftex]{hyperref}
\idline{75}{282}
\begin{document}
\def\teff{$T\rm_{eff }$}
\def\kms{$\mathrm {km s}^{-1}$}
\def\ha{H$\alpha$}

\title{
Mapping the Universe with slitless spectroscopy
}
   \subtitle{}

\author{
P. \,Monaco\inst{1,2,3,4}, on behalf of the Euclid Consortium
}

\institute{
Universit\`a di Trieste, Dipartimento di Fisica, Via Tiepolo 11, I-34131 Trieste, Italy
\and
Istituto Nazionale di Astrofisica -- OATs, Via Tiepolo 11,
I-34131 Trieste, Italy
\and
Istituto Nazionale di Fisica Nucleare Trieste, via Valerio 2, I-34127 Trieste
\and
Institute for the Fundamental Physics of the Universe, via Beirut 2,
I-34151 Trieste\\
\email{pierluigi.monaco@inaf.it}
}

\authorrunning{Monaco }

\titlerunning{\Euclid's slitless spectroscopy}

\date{Received: Day Month Year; Accepted: Day Month Year}

\abstract{
\Euclid will survey most of the accessible extragalactic sky with imaging and
slitless spectroscopy observations, creating a unique spectroscopic catalog of
galaxies with {\ha} line in emission that will map the Universe from $z=0.9$ to
$1.8$. With low expected statistical errors, the error budget will likely be
dominated by systematic errors related to uncertainties in the data and
modelling. I will discuss the strategy that has been proposed to mitigate the
expected systematic effects and propagate the uncertainty of mitigation to
cosmological parameter errobars.
}
\maketitle{}

\section{Introduction}

Observations of the Cosmic Microwave Background (CMB, Planck Collaboration 2018) 
have provided percent-accurate constraints to cosmological parameters,
strengthening the case for a 6-parameter flat $\Lambda$CDM cosmological model;
this is consistent with most available evidence on large scales, but at the cost
of leaving 95\% of the present mass-energy budget unexplained. In fact, most
matter today is thought to be in the form of an unknown collisionless particle,
and most energy today is thought to be in a dark energy component that is
accelerating the Universe expansion, represented by a positive cosmological
constant.

To shed light on the dark sector, it is crucial to map the Universe at lower
redshift, when dark energy becomes dominant. The European Space Agency has
promoted the \Euclid mission (Laureijs et al. 2011), 
an optical and near-infrared
space telescope that will survey the sky with a visual imager (VIS), optimised
for galaxy lensing, and a near-infrared imager and spectrograph (NISP),
optimised for galaxy clustering. 

\section{Galaxy Clustering with slitless spectroscopy}

\begin{figure*}[t!]
\centering{
\resizebox{0.56\hsize}{!}{\includegraphics[clip=true]{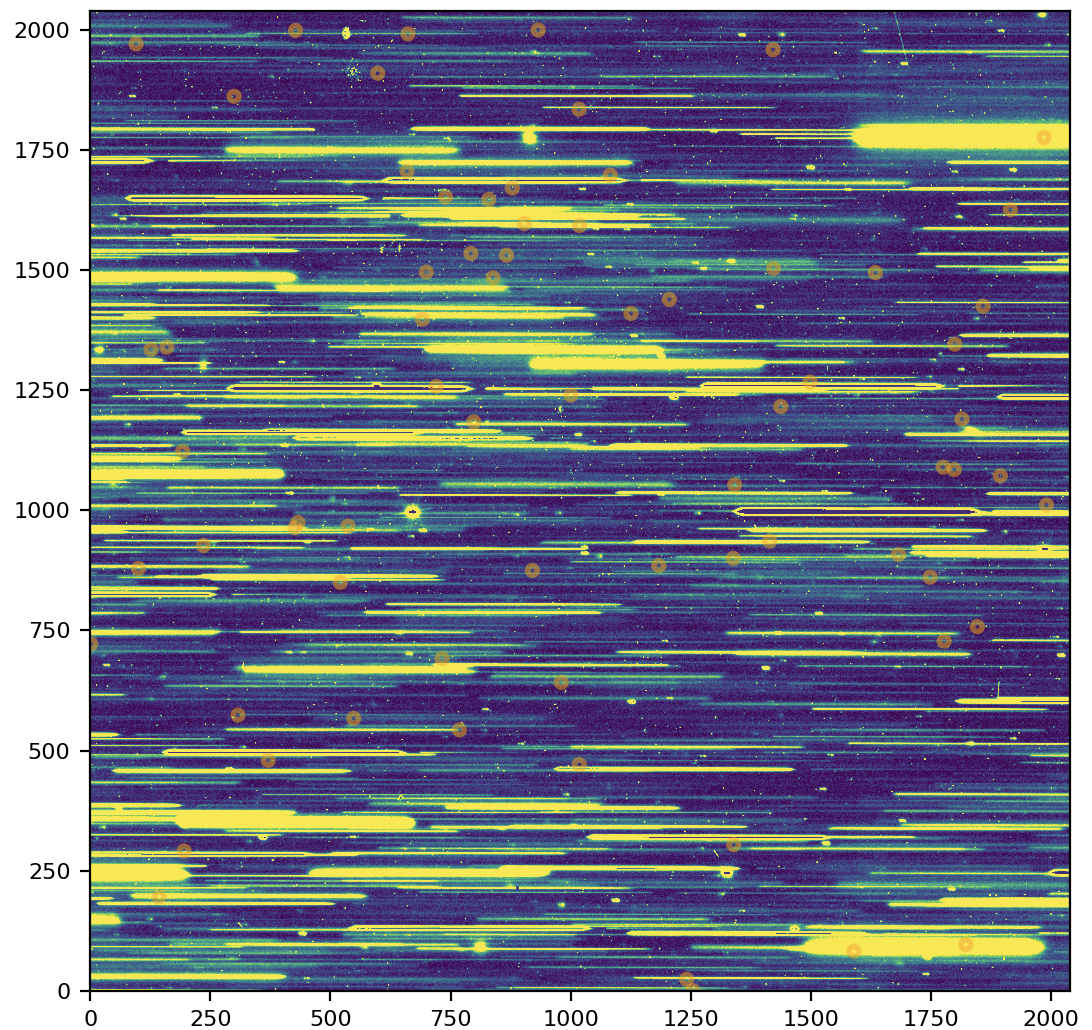}}}
\caption{\footnotesize Simulations of a NISP exposure, where each source
  produces a stripe along the dispersion imprinted by the grism.
  Circles denote the positions of the {\ha} lines. Credit: Ben Granett, OU-SIM an OU-SIR teams.
}
\label{slitless}
\end{figure*}

Spectroscopic observations from space are complicated by the impossibility to
use screens with suitably pierced holes or slits, as customary from the ground.
While JWST has implemented a novel micro-mirror technology that provides a way
to select what part of the image is to be dispersed by a grism, \Euclid will
adopt a straight slitless spectroscopy strategy, already experimented with
Hubble Space Telescope (Bagley et al. 2020).
This means that each source will produce a
straight track on the NISP detector, as showed in the simulation in
Fig.~\ref{slitless}. These tracks will be separated and used to produce  1D
spectra with a resolution of $R=380$. This observing strategy will allow us to
detect emission-line galaxies (ELGs) and measure their redshift, provided that
the emission line is correctly recognised. The grism will be sensitive to
wavelengths in the range $\lambda \in [1.25,1.85]$ {\micron}, so the most
abundant detectable ELGs are expected to be {\ha} emitters in the redshift range
$z\in[0.9,1.8]$. Simulations show that the probability of detecting an ELG drops
for line fluxes below $2\times 10^{-16}$ erg s$^{-1}$ cm$^{-2}$, the nominal
line-flux limit of the Euclid Wide Survey.

The obvious drawback of this strategy, the need to deblend all the sources in a
field, is balanced by the ability to perform a matter-of-fact blind search of
emission-line galaxies. Indeed, spectra will be extracted for any source detected in the photometric observations, that we may think as limited to $\HE<24$, and the number of
{\ha} ELGs associated to fainter sources has been demonstrated to be negligible
in Bagley et al. (2020). 
At the same time, although we expect the success rate of
deblending to be a function of surface density of all sources, the sample will
be free of fiber collision bias (e.g., Bianchi et al. 2018).

\section{Chasing systematic effects}

The Observational Systematics Work Package of the Galaxy Clustering Science
Working Group has surveyed the whole pipeline, from raw data to the measure of
galaxy clustering that is provided to the likelihood. Using the formalism of
Monaco, Di Dio \& Sefusatti (2019), 
we have classified the possible systematic effects as follows:
(i) modulations of the effective flux limit of the sample, due either to
instrumental issues (e.g. the tiling of the various dithers will produce an
inhomogeneous exposure time map, while straylight from nearby bright stars will
modulate the noise) or to astrophysical foregrounds (e.g. zodiacal light will
add to the background noise, while Milky Way extinction will decrease the
signal); (ii) redshift errors due to line misidentifications; (iii) redshift
errors due to noise fluctuations being interpreted as lines in an overall
undetected spectrum, thus creating `noise interlopers'.

The first class of systematic effects will be mitigated by suitably constructing
a random catalog. To get rid of the angular footprint of a survey, clustering
estimators usually compare the density field measured with the data sample with
that from a random sample that covers the same area and is unclustered on the
sky; its number density is usually taken to be 50 times that of the data sample (fitted by a model in order not to erase some radial modes),
to minimise the amount of extra shot noise introduced by the random. We will
construct the random by forward-modeling the completeness and purity of the
spectroscopic sample, thus creating what we call a visibility mask. This will be
done by taking profit of the Euclid Deep Field, where 50 deg$^2$ of the sky
will be surveyed ten times with various orientations of the grism, and with 40
more pointings with a `blue grism' that is sensitive in the range $\lambda \in
[0.92, 1.25]$ \micron; from it, we will extract a bona fide sample, pure at $\sim99$\% level, of
the ELGs that can be seen in the EWS. These galaxies will
be used to create a parent random catalog that is unclustered on the sky, whose objects
have the same physical properties of the target sample; these random galaxies
will be injected in the EWS NISP images, and processed to determine their
probability of detection. This way the space density of the selected random will
be modulated on the sky by systematics in the same way as the data sample.

While the third class of systematic effects, the noise interlopers, can be
modeled by suitably adding a class of contaminants to the random catalog, the
second class, line misidentifications, is more subtle to address (Addison et al. 2019).
With a typically steep luminosity function of sources, most
objects in a catalog are around the detection limit, where a single emission
line is typically detected. If no further information is used, the contamination
level is expected to be around $\sim 10$--20 \%, mostly coming from $[$O{\sc iii}$]$
emitters at higher redshift. These galaxies will be moved from their redshift to
the one corresponding to {\ha}, carrying with them their rescaled clustering
signal, so the measured two-point function $\xi(r)$ will be the weighted sum of
the target one $(1-f)^2 \xi_{\rm target}$ and the contaminant one $f^2 \xi_{\rm
  interloper}$, where $f$ is the fraction of contaminants in the sample. This
contamination can be mitigated at the likelihood level, comparing the
measurements with a weighted sum of predictions relative to the target sample
and to the significant contaminants, with $f_i$ fractions treated as nuisance
parameters subject to a tight prior coming from measurement of the Deep Field.

\section{Propagating the uncertainty in the mitigation}

This mitigation strategy will anyway leave residuals that contaminate the
sample. Every step in the modeling of the visibility mask of the EWS has an
associated uncertainty that must be propagated to parameter errorbars. The most
effective way to achieve this is to construct a set of simulated mock galaxy
catalogs and process them in the same way as the parent random catalog described
above: inject galaxies in the images and compute their probability of being
detected. However, this process should not be performed using our best knowledge
of the visibility mask but a modulation of it, obtained by perturbing every
single step in the pipeline, sampling its estimated error PDF. As an example,
detection probability will depend on the measured noise level, and we will use
the best-fit value of the noise to create the random, and a value drawn from its
PDF for applying the visibility mask to mock galaxy catalogs.

\begin{figure*}[t!]
\resizebox{0.9\hsize}{!}{\includegraphics[clip=true]{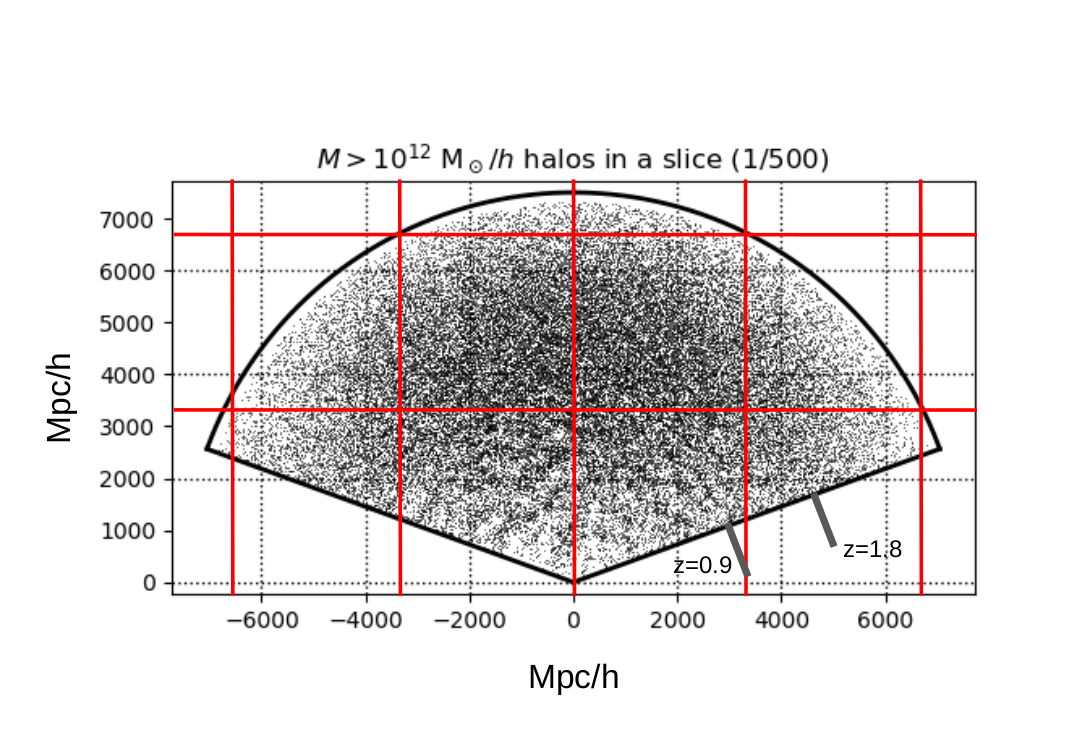}}
\caption{\footnotesize Lightcone generated with \texttt{PINOCCHIO}, see text for explanations.
}
\label{mocks}
\end{figure*}

This approach will provide a brute-force numerical estimate of the covariance
matrix that will include both cosmic covariance and the uncertainty in the
mitigation of systematic effects. Proper sampling of the matrix requires
thousands of mocks; in this moment the Galaxy Clustering Science Working Group
is preparing 3500 simulations of the \Euclid sky. N-body codes are simply too
expensive to address this massive production, so we will resort to approximate
methods (Monaco 2016). 
I am presently working to prepare such a large
set of simulations using the \texttt{PINOCCHIO} code (Monaco, Theuns \& Taffoni 2002; Munari et al. 2017),
based on Lagrangian Perturbation Theory. Fig.~\ref{mocks} shows one of these
lighcones, reporting dark matter halos (with $M_h>10^{12}\ M_\odot$, one in 500)
in comoving coordinates, in a slice that cuts through the survey volume; the
catalogs cover half of the sky, with the exception of unobserved low Galactic
latitudes, and start at $z=4$. The red lines mark the box size (3380 $h^{-1}$
Mpc), that is tiled to cover the survey volume. When ready, this will be the
largest set of cosmological simulations ever produced.

\begin{acknowledgements}
P.M. thanks L. Guzzo, W. Percival, Y. Wang, C. Scarlata, B. Granett, M. Moresco,
S. De La Torre and the members of the Observational Systematics Work Package for
many discussions. The simulation used for Fig. 1 was produced by the Euclid
Science Ground Segment, in particular Operational Units OU-SIM and OU-SIR.
\AckEC
\end{acknowledgements}

\end{document}